# Atom-by-atom construction of attractors in a tunable finite size spin array

A. Kolmus[1], M.I. Katsnelson[2], A.A. Khajetoorians[2], and H.J. Kappen[1]

[1]*Donders Institute for Neuroscience, Nijmegen, The Netherlands*

[2] *Institute for Molecules and Materials, Nijmegen, The Netherlands*

**Abstract**

We demonstrate that a two-dimensional finite and periodic array of Ising spins coupled via RKKY-like exchange can exhibit tunable magnetic states ranging from three distinct magnetic regimes: (1) a conventional ferromagnetic regime, (2) a glass-like regime, and (3) a new multi-well regime. These magnetic regimes can be tuned by one gate-like parameter, namely the ratio between the lattice constant and the oscillating interaction wavelength. We characterize the various magnetic regimes, quantifying the distribution of low energy states, aging relaxation dynamics, and scaling behavior. The glassy and multi-well behavior results from the competing character of the oscillating long-range exchange interactions. The multi-well structure features multiple attractors, each with a sizable basin of attraction. This may open the possible application of such atomic arrays as associative memories.

## 1. Introduction

Due to the growing demand for energy-efficient information and computing technologies [1] as well as, artificial neural networks, there has been vast interest in the development of hardware designed for efficient pattern recognition tasks [2-6]. The strategy toward this end has been to create physical analogues of machine learning concepts in materials [7]. Manipulating the spin degree of freedom in solid-state matter is a promising route for brain-inspired computing [8, 9], due to the combination of (i) high-quality materials available to which individual and coupled moments can be manipulated down to the atomic scale, (ii) rich landscape of non-linear, dynamic, and stochastic spin-based phenomena, (iii) the variety of read/write options available. Recently, many schemes have been proposed which utilize the spin degree of freedom in hardware, to perform machine learning tasks [10-13]. Of particular interest is the Hopfield network: a recurrent neural network that implements an associative memory [14]. The neurons are binary variables that evolve under a stochastic (Glauber) dynamics [14]. A given memory pattern is stored as a local energy minimum, or so-called attractor, in a tailored energy landscape comprised of many local minima. When the system is initialized in a distorted version of one of its stored memories, the network state evolves to the nearest attractor and thus restores the memory as a form of pattern recognition. The memory is called associative, because a single memory can contain many components and one component can restore the entire memory. A Hopfield network is formally equivalent to an array of coupled Ising spins, where the magnetization of each spin represents a pixel and the couplings are optimized to store a given set of memories [14].

A key challenge toward a material realization of the Hopfield network is creating many low-energy complex and tailored ordered states, to identically match a desired pattern. Conventional magnetically ordered states such as ferromagnets exemplify the concept of broken symmetry [15] with bistability of the ground state. However, this bistability is not ideal for attractor memory, because at most one pattern (and its inverse) can be stored. At the opposite extreme is the spin glass [16-18], as suggested by Edwards and Anderson in 1975 [19, 20], that is characterized by an energy landscape with infinitely many local energy minima separated by energy barriers of multiple heights so that there is a broad distribution of transition times between different minima. However, spin glasses cannot be used for associative memory either, because the basins of attraction are too small. Therefore, an intermediate potential landscape, situated between a too simple and robust double-well landscape and a too complex glassy landscape is necessary to realize the Hopfield network in real magnetic materials.

Here, we propose that an ordered and finite 2D array of spins interacting via a well-defined long-range RKKY interaction [21] can be used to create energy landscapes with a tunable level of complexity. By changing the ratio α= λ/a of the RKKY wavelength (λ) and the lattice constant (a), the energy landscape of the spin array can be tuned between three different magnetic regimes, ranging from a regime with double-well potential (DW), through a multi-well potential (MW) regime, to a spin glass-like regime, similar to what has been previously studied in ref. [22, 23]. We refer to this regime as a spin Q glass (SQG) (Figure 1) [24]. To characterize each regime, we compute the distribution of the spatial Fourier components (Q-space) of the low energy states allowing us to subsequently classify each regime and construct a regime diagram. Subsequently, we characterize each regime by computing the scaling of the number of local minima with system size; we use a waiting time analysis to characterize the aging dynamics [25]; and we estimate the size of the basins of attraction in each of the regimes.

2. Methods and Results

   2.1 Modeling the magnetic interactions in the Ising spin array

We consider simulations of $n \times n$ 2D rectangular spin arrays ($4 \leq n \leq 40$, unless specified otherwise $n = 25$) with interatomic distance $a$, as shown in Fig 1a. We consider an exchange Hamiltonian of the form:

$$H = -\sum_{i>j} J_{ij} s_i s_j \tag{1}$$

where $s_i$ represent an Ising spin with the position $i$. The exchange parameter $J_{ij}$ is derived from an isotropic RKKY-like exchange [26]:

$$J_{ij} = \begin{cases} 0 & , i = j \\ \frac{1}{r_{ij}^2} \sin\left(\frac{2\pi}{\lambda} r_{ij}\right), & i \neq j \end{cases} \tag{2}$$

where $r_{ij}$ is the distance between positions $i$ and $j$, and $\lambda$ is the period of the RKKY interaction as illustrated for the center atom in Fig. 1a. As we show, the competition between the lattice constant and the resultant exchange interactions leads to three distinguishable regimes, characterized by different distribution and density of low-energy ground states (Fig. 1b).

### 2.2 Q-space and energy histograms as regime identifiers

We analyze the energy landscapes for different $\alpha$ and $n$ in the following way. We use iterative improvement, which is a zero-temperature single spin flip dynamics, to find metastable states that are fixed points of the spin dynamics. We initialize a state randomly, run the dynamics and repeat this procedure 5000 times. We characterize the empirical distribution of low energy states thus obtained by computing the Q-space histogram, or Q-histogram for short. The Q-histogram is the sum of the absolute values of the Fourier coefficients of the metastable states, weighted by the empirical probability of finding the state. As we illustrate in a schematic in Fig. 1c, we will subsequently show that there are distinctly different Q-histograms, which we can identify with the magnetic regimes illustrated in Fig. 1.

We vary the ratio $\alpha$ from 1 to 50 with a step size of 0.5, and analyze the resultant properties of the array. In Fig. 2a, the Q-histogram for different $\alpha$ for a $25 \times 25$ lattice is shown. For $\alpha$ = 2.5, 7.5-12, 46 the array is in the SQG, MW, and DW regime, respectively. The value $\alpha$ = 4 marks the transition points between SQG and the MW regime, which is distinguished by the emergence of a larger number of states centered around Q = 0. The transition between the MW and DW regime is more continuous and happens around $\alpha \sim 40$, in which only a strong global maximum centered around Q = 0 remains and all meta stable states disappear (Fig. 2b). Furthermore, we distinguish two MW regimes, *a* and *b*. The difference between MW-a and the SQG regime is that in the former the ground state is an ordered state whereas in the latter it is a disordered state. The difference between MW-a and MW-b is that in MW-a, the ground state probability is vanishingly small; while in MW-b, the probability is significant (Fig. 2 b). As we show later, the MW-b regime is the regime most suitable for attractor memory.

For $\alpha$ = 2.5, the Q-histogram is characterized by four peaks, at non-zero Q values, each along either the negative/positive $Q_x$ or $Q_y$ axis. The corresponding picture in real space, as illustrated in in Fig. 3, is that there are many low lying metastable states with no clear long-range ordered pattern, similar to previously observed self-induced glassiness [22, 23]. Therefore, we identify this regime as the SQG regime. As $\alpha$ is increased, the diameter of the ring in the Q-histogram shrinks and the center starts to gain intensity with the state Q = 0, being the global maximum. This results in a preferable ferromagnetic ground state, but co-exists with a distribution of low-energy meta-stable states with non-trivial Q values. Therefore, we identify this as the MW regime. This leads to a distribution of states, as shown in real space, which cannot be characterized by simple ferromagnetic or antiferromagnetic order, as shown in Fig. 3. As $\alpha$ is increased, the distribution of states around Q = 0 narrows, leading to fewer metastable states and the growing dominance of the ferromagnetic ground state. This coincides with the increasing wavelength to the RKKY, relative to the total lattice size, leading to the dominance of one sign of the RKKY exchange (Fig. 1a). For $\alpha > 40.00$ we transition from the MW regime into the DW regime, also leading to a clear ferromagnetic pattern in real space (Fig. 3). This is based on the behavior of the system, see the discussion of Figure 3.

In Fig. 2b, histograms of the state energies are shown for various values of $\alpha$ corresponding to the different regimes. The histogram consists of fifty equally spaced intervals between the ground state energy $E_0$ and the highest energy indicated in the bottom right corner of each graph. The total area of the histogram is normalized to one. The energy histograms are in agreement with what one expects for the energy landscapes for the different regimes, as illustrated in Fig. 1b. The SQG shows a Gaussian like distribution ($\alpha = 2.50$), indicating the presence of many meta-stable states over a narrow range of energies. In direct contrast, for $\alpha = 46.00$, there is strong intensity near the ground state indicative of a strong double-well potential. In the MW regime, there are two distinctly different distributions depending on the value of $\alpha$. For $\alpha = 7.50$ (MW-a regime), there are many low energy states, similar to the SQG regime, even though the ground state resides at Q = 0 (Fig. 2a). As $\alpha$ increases, the gap in energy between the ground state and the other low energy states grows. This leads to a second type of behavior in the MW regime, exemplified by $\alpha = 12$ (MW-b regime). There are fewer low energy states in comparison to the MW-a regime, and the ferromagnetic ground state is more dominant in comparison. The MW-b ($\alpha = 27$) lattice resembles the stable retrieval phase A [27] for the Hopfield model where stable attractors (see later) are the lowest energy states, while the MW-b ($\alpha = 12$) resembles retrieval phase B [27], where stable and unstable attractors have similar energies. As we will show, the MW-b regime has the right balance of a large number of low energy states and large basins of attraction for potential application to a Hopfield network.

**2.3 Scaling and entropy based behavior – constructing the regime diagram**

Utilizing the aforementioned distinctions between regimes, we construct a regime diagram delineating the different magnetic regimes (Fig. 4a). The horizontal axis represents α, from 2.5 to 50 with a step size of 0.5. The vertical axis represents the number of spins along one axis of the square lattice, from *n* = 4 to *n* = 40 spins. For each grid-point $(n, \alpha)$ statistics have been obtained by 5000 iterative improvement runs. From these statistics, we calculated the entropy $S$ over the states, $S = -\sum_i p_i \ln p_i$ where $p_i$ is the probability of finding a unique state $i$. The entropy is a measure of the roughness of the energy landscape, as landscapes with many minima have higher entropy compared to landscapes with fewer minima. The entropy for each grid cell is indicated by the intensity defined in the scale bar. We observe a strong enhancement of the entropy as the value of α is smoothly decreased. The increased entropy is due to the increasing number of metastable states seen in the Q-histograms, discussed in Fig. 2a. We note that the entropy is significantly underestimated in the SQG and MW-a regimes, because it is virtually impossible to sufficiently sample to get a correct estimate of the entropy due to the large number of available states. The regime boundary between the SQG and MW is based on Q-histograms, namely if the Q = 0 state is the ground state as discussed earlier. This method does not work well for distinguishing the MW and DW regimes, since the ferromagnetic ground state is very strong along the regime boundary. We found empirically that an entropy value below 0.95 is indicative of DW behavior and used this to create the regime boundary. Nevertheless, the entropy behavior illustrates that the various regimes we see here cannot be ascribed to an effect of small finite sized systems, and that there are large regions where one can find MW-b regimes which, as we described later, can be used as attractors.

While the value of entropy is a strong indicator of the different regimes, it is interesting also to investigate the scaling behavior of the number of low lying states with system size in the different regimes. For instance, we expect a strong difference in the scaling behavior between a glass like system and a ferromagnetic system. Since the boundaries depend on both *n* and α, we compute the number of meta stable states along the white dashed lines in Fig. 4a. These lines were chosen such that they followed regions of similar entropy / energy landscapes and thus show the scaling behavior for lattices of difference with similar dynamics. The results in Fig 3b show exponential scaling in the SQG regime. To measure the exponential scaling of the SQG regime, we used 1000000 initializations instead of 5000. However even for $n$ = 11 this is not sufficient, illustrating the exponential growth. The results in Fig. 4b show polynomial scaling in the MW-b regime and linear scaling in the MW-a regime. As expected, the number of metastable states does not grow with system size in the DW regime.

### 2.4 Aging dynamics

In order to explore the relaxation dynamics of the various regimes, we perform aging calculations similar to ref. [25]. We consider a randomly initialized array and let this array evolve over time using Metropolis-Hastings algorithm [28, 29] with a suitable, fixed, temperature *T*. The aging dynamics are captured by calculating the autocorrelation function between the state at $t = t_w$ and a later time $t$:

$$C(t_w + t, t_w) = \frac{1}{N} \sum_i s_i(t_w) * s_i(t_w + t) \tag{3}$$

where $N$ is the number of spins. Increasing $t_w$ increases the probability that the system has reached a favorable low-energy state, resulting in larger autocorrelation. A constant value of $C(t_w + t, t_w)$ for large $t_w$, indicates that the array has relaxed into a frozen state. Since $C(t_w + t, t_w)$ at a given temperature depends on the smoothness or ruggedness of the energy landscape, we can use it to analyze the different regimes.

Fig. 5a shows $C(t_w + t, t_w)$ for each regime discussed in Fig. 2a, for α = {2.50, 7.50, 27.00 and 46.00} and $t_w = \{2^5, 2^7, 2^9, 2^{11}, 2^{13}, 2^{15}, 2^{17}\}$. As expected, for the DW regime, there is only one relaxation time-scale independent of $t_w$ ($t$ = 10³), but the asymptotic correlations $C(t_w + t, t_w)$ depends on the value of $t_w$. These are the characteristics of a ferromagnetic system, where for large $t_w$ the system cannot escape from one of its bistable modes. The SQG regime shows the other extreme, where the asymptotic correlations are zero for all $t_w$ but with multiple relaxation times. This confirms that the SQG regime exhibits aging properties of a spin glass, as was previously considered seen for the prototypical spin glass Mn-Cu [25]. The MW regime shows the intermediate case between the SQG and the DW; both the relaxation times and the asymptotic value depend on $t_w$. Most notably, $C(t_w + t, t_w)$ exhibits plateaus, not found in the SQG and DW regimes. The system initially relaxes towards a relatively stable meta stable state and at a much larger time scale relaxes to the ground state. This is especially true for α = 27.00, where two plateaus can be clearly distinguished due to the large difference between the two relaxation times. This particularly observation clearly illustrates the intermediate behavior of the MW regime, in which multiple yet robust energy minima are present, as to which the system can relax. Moreover, in combination with the Q-histogram analysis and density of states analysis, this aging analysis is a quantitative measure for distinguishing the various regimes.

### 2.5 Basins of attraction for each magnetic regime

We have demonstrated that we can modify the energy landscape of the spin array, by changing the gate parameter α, distinguishing three different regimes characterized by different distribution of energy states, Q-states, as well as aging dynamics. With regards to applying these systems to create associative memories, it is important to characterize the robustness of the minima in these different regimes. In Fig. 6a, we explore the robustness of the low-energy states by studying the basin of attraction. We estimate the return probability of the 20 lowest energy states by randomly flipping a percentage of the spins of the given low-energy state and running the zero temperature dynamics until convergence, and measuring the return probability to the original state. This return probability is the average over 200 runs per state for an initial perturbation = {0.05, 0.10, 0.15, 0.20, 0.25, 0.30} %. We plot this return probability R for each regime for different perturbation percentages, where the colors correspond with the energy of the studied state. The SQG regime has a zero percent return probability for all states even when the initial perturbation affects only 5% of the lattice, as expected for a glassy landscape. The strongest state in the MW-a regime has a return probability of 40% for a 5% initial perturbation, which is not robust enough to serve as associative memory. In contrast, the MW-b shows many robust memories. We define stable attractors as metastable states that for an initial perturbation of 10 % have a return probability to that state of 0.9 or higher. The results in Fig. 6a show that the MW-b regime has 32 stable states. In the DW regime, only the ferromagnetic ground state has a large basin of attraction, while all meta-stable states have very weak basins of attraction.

In Fig. 6b, we determined the number of stable states for lattices of size 10, 20 and 25. We varied $\alpha$ in the range where the lattice is in the MW-b regime. For each $\alpha$, we determine the stability of the states with 400 lowest energy levels by repeating the above-mentioned procedure. In Fig. 6c and 6d, we illustrate examples of the coinciding real space magnetization distribution of stable and unstable states, respectively, and compare the average states before and after the return procedure. Stable patterns appear to be more regular and their periodicity is similar to the periodicity of the RKKY interaction. When perturbing a stable pattern by 10 % it robustly returns to the original state more than 90 % of the time (Fig. 6c). When perturbing an unstable pattern by 10 % it does not return to the original state (Fig. 6d).

3. Conclusion

In conclusion, we have shown that finite size spin lattices with long-range competing RKKY interactions can serve as a platform to create a rich variety of magnetic regimes, ranging from robust double well potentials, toward glassy landscapes and multi-well landscapes. The necessity of such states are recently discussed in the context of biological complexity [30, 31]. In addition to the peculiarity that we see spin glass like behavior in relatively small systems, the multi-well landscapes are at the edge of chaos, between the ferromagnetic DW regime that is too simple and the spin glass regime that is too complex. The polynomial growth of the number of local minima with respect to the system size can be considered as a characteristic feature of the multi-well regime, which has not been previously studied. Moreover, as we show, the multi-well regime can be utilized to create stable attractors in spin-based hardware, serving as a platform for associative memory and recurrent neural networks. Importantly, these systems can be created, based on recent experiments where the RKKY interaction between individual magnetic atoms can be tuned by atomic-scale manipulation [32, 33]. As a further point of study, it will be interesting to go beyond the Ising limit, and quantify how the presence of long-range anisotropic exchange interactions, like non-collinear orientations of spins seen in experiments, can modify the proposed regimes seen here in the Ising limit [34, 35].

A.A.K. acknowledges funding from NWO, and the VIDI project: "Manipulating the interplay between superconductivity and chiral magnetism at the single-atom level" with project number 680-47-534. A.A.K. also acknowledges funding from the European Research Council (ERC) under the European Union's Horizon 2020 research and innovation programme (SPINAPSE: grant agreement No 818399). H.J.K. acknowledges funding by ONR Grant N00014-17-1-256.

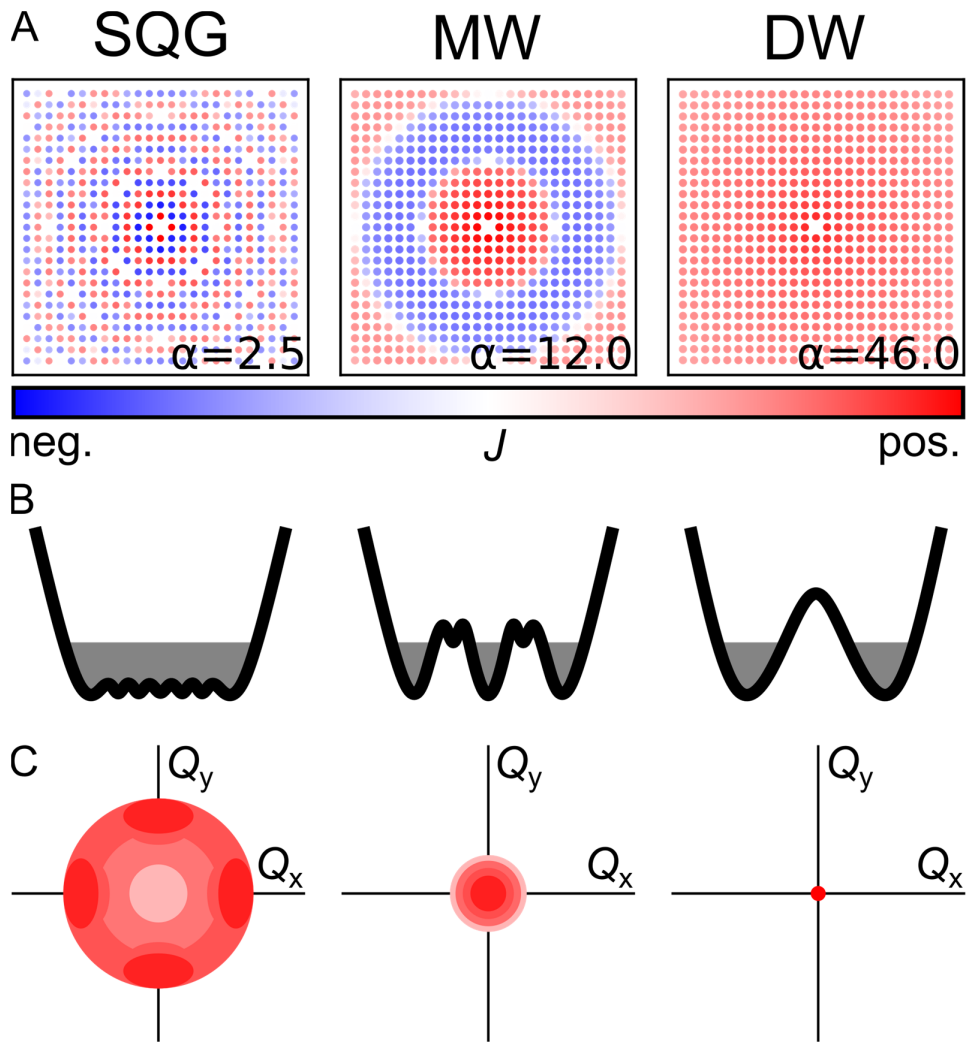

**Figure 1:** (a) The spatial distribution of the RKKY exchange interaction ($J$) for the central atom in the Ising spin array ($n$ = 25) for different α for the labeled magnetic regime: spin-Q glass, multi-well, double well. The color bar represents the amplitude and sign of the interaction. (b) Schematic of the energy landscape for the three-labeled regimes, illustrating qualitatively the distribution and depth of states for each regime where grey illustrates the effective temperature. (c) Illustration of the distinguishing features in the Q-space histogram identifying each regime, where white to red intensity corresponds to low to high number of states.

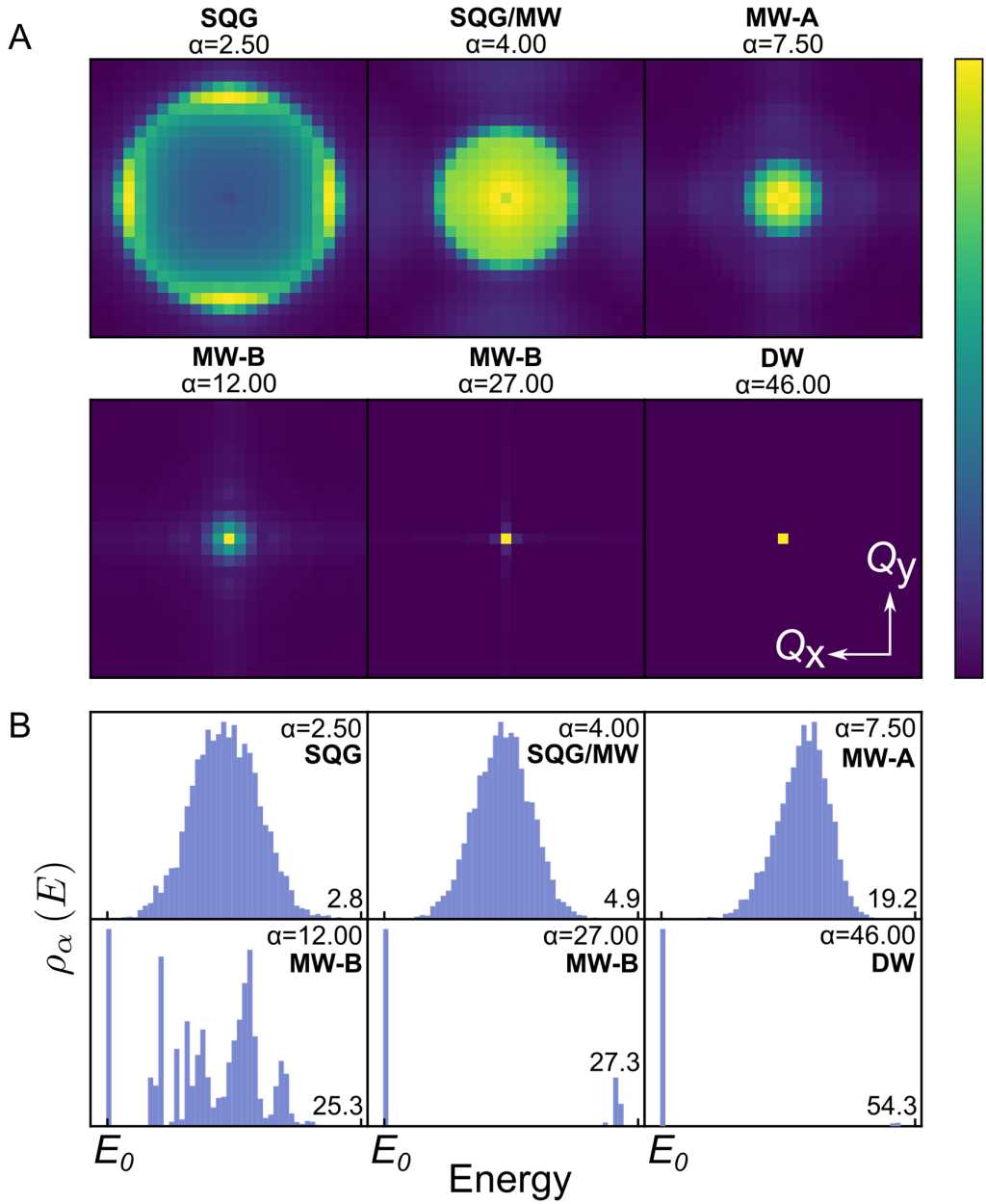

**Figure 2:** (a) Q-histogram for various values of α. The histograms are the weighted average over the absolute values from the Fourier transform of the metastable states. The color bar represents the total number of states. (b) Histograms of the distribution of energy states for each α shown in (a), where the x-axis scales from the ground state ($E_0$) to the maximal state. The total number of bins in each histogram corresponds to 50. The distribution is normalized by the total area of the distribution. The density of states ($\rho_\alpha(E)$), as defined in the text for the indicated values of energy.

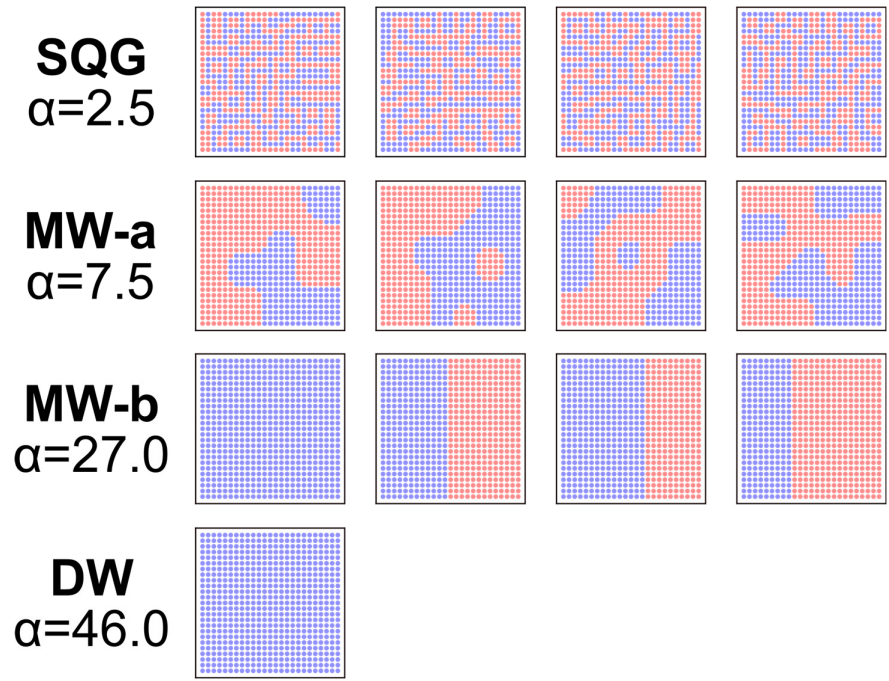

**Figure 3**: Examples of the real space magnetization distribution for a lattice size of 25 x 25, for various metastable states for each of the labeled regimes (red/blue correspond to an average spin value of -/+ 1). Each of the patterns corresponds to a low-energy state, taken from the histogram in Fig. 2b, for the labeled regime and value of α. The states go up in energy from left to right.

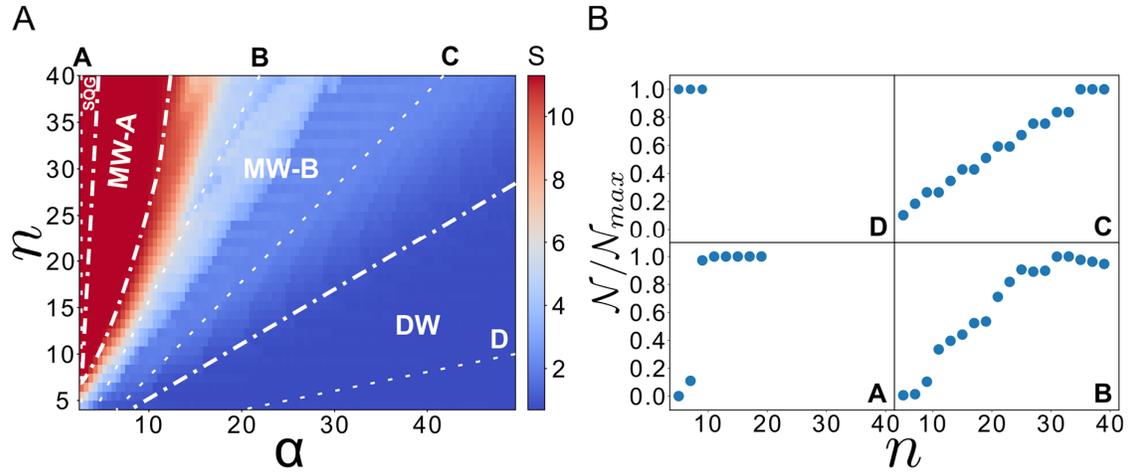

**Figure 4:** (a) A regime diagram with the lattice width *n* on the vertical axis and α on the horizontal axis. The white dashed lines indicate the different regimes, as labeled and as defined by the corresponding Q-histograms. The color scale indicates the entropy as defined in the main text. (b) The scaling behavior of near the boundary between each regime, corresponding to the white dashed lines in (a) and labeled by the letters A-D. Each plot corresponds to the number of available states as a function of the lattice width *n* vs the normalized number of metastable states $N/N_{max}$. $N$, $N_{max}$ are the number of stable states and the largest number of stable states per graph, respectively, and we divide the former by the latter in order to normalize each plot for comparison.

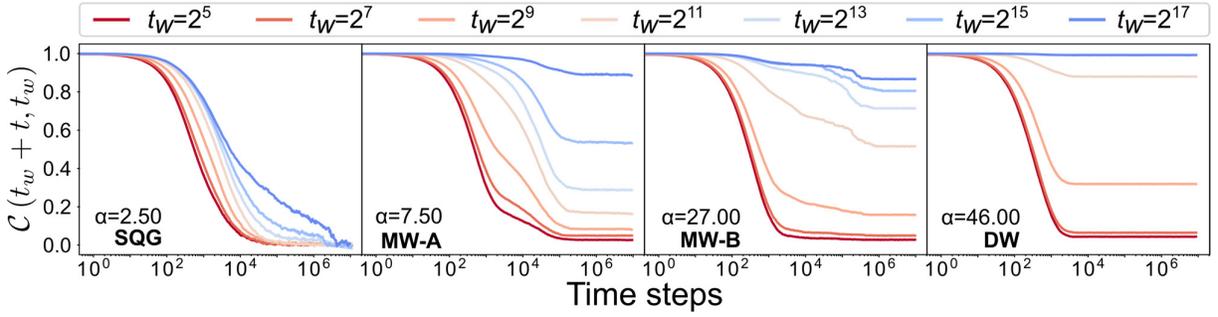

**Figure 5:** The autocorrelation function $C(t_w + t, t_w)$, as defined in the text, for different $\alpha$ and labeled regimes, where $t_w$ is the waiting time before measuring the autocorrelation and $t_w$ is the time step during the measurement as indicated by the colors/values labeled above the graphs. Each line is the average over 100 runs. For each α the temperature was set below the critical temperature (determined using the Binder cumulant), but high enough to show aging behavior in $10^7$ time steps.

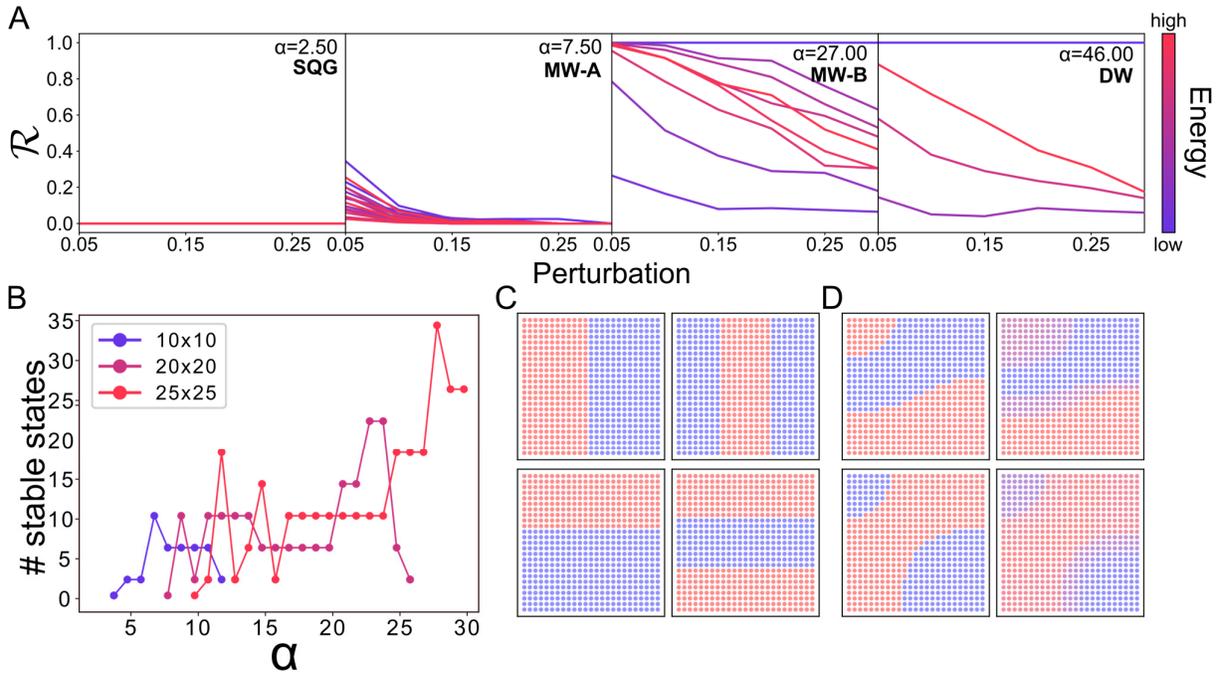

**Figure 6**: (a). The return probability $R$ for a lattice of size 25 x 25 of the 20 lowest energy states as a function of the initial perturbation, where perturbation is defined as the percentage change of spin flips. Each line represents a single state, with low energy states having a purple hue and higher energies shifting towards a red hue, as labeled by the color bar on the right. The return probability per state was calculated over 200 individual runs per perturbation percentage. (b) The number of stable states, as defined in the text, as a function of $\alpha$ for different labeled array sizes. (c) Examples of the magnetization distribution for various stable attractors in the MW-b regime. The color indicates the average spin value from minus one (red) to positive one (blue). (d) The evolution of unstable attractors in the MW-B regime before the perturbation (left side) and after the return procedure (right).